# Traffic jams and driver behavior archetypes

Shawn Berry, DBA[1*]

October 5, 2025

[1]William Howard Taft University, Lakewood CO, USA

*Correspondence: shawnpberry@gmail.com

**Abstract**

Traffic congestion represents a complex urban phenomenon that has been the subject of extensive research employing various modeling techniques grounded in the principles of physics and molecular theory. Although factors such as road design, accidents, weather conditions, and construction activities contribute to traffic congestion, driver behavior and decision-making are primary determinants of traffic flow efficiency. This study introduces a driver behavior archetype model that quantifies the relationship between individual driver behavior and system-level traffic outcomes through game-theoretic modeling and simulation (N = 500,000) of a three-lane roadway. Mann-Whitney U tests revealed statistically significant differences across all utility measures ($p < .001$, $d > 2.0$). In homogeneous populations, responsible drivers achieved substantially higher expected utility (M = -0.090) than irresponsible drivers (M = -1.470). However, in mixed environments (50/50), irresponsible drivers paradoxically outperformed responsible drivers (M = 0.128 vs. M = -0.127), illustrating a social dilemma wherein defection exploits cooperation. Pairwise comparisons across the six driver archetypes indicated that all irresponsible types achieved equivalent utilities while consistently surpassing responsible drivers. Lane-specific analyses revealed differential capacity patterns, with lane 1 exhibiting a more pronounced cumulative utility decline. These findings offer a robust framework for traffic management interventions, congestion prediction, and policy design that aligns individual incentives with collective efficiency. Directions for future research were also proposed.

## 1. Introduction

### 1.1 Background and Rationale

Traffic congestion is a persistent and costly issue faced by modern transportation systems worldwide (Bhardwaj, 2023). The economic impact of traffic jams extends beyond individual travel delays, encompassing broader societal costs, such as increased fuel consumption, environmental harm, and productivity losses. As urbanization and transportation demand increase, understanding the mechanisms behind traffic jam formation has become increasingly important (Habiba, 2023). This is a complex system in which individual driver decisions interact with the outcomes of the collective system (Habiba & Talukdar, 2023). Traditional traffic engineering has provided insights into flow dynamics but has often overlooked the behavioral and strategic aspects of driver decision-making (Li, 2023). Interdisciplinary research indicates that insights from behavioral economics, game theory, and decision science can enhance our understanding of traffic phenomena. Intelligent transport systems and autonomous vehicle technologies offer new opportunities to study and influence traffic patterns (Munigety 2023). However, their successful implementation relies on a thorough understanding of human driver behavior within traffic systems (Munigety and Mathew, 2023). This is particularly crucial during the transition phase, when human-driven and autonomous vehicles will coexist on roads (Yu, 2023). Game theory has been utilized to examine strategic interactions in traffic systems in which individual decisions impact collective outcomes (Yu & Zhao, 2023). The Nash equilibrium highlights stable traffic patterns that emerge from strategic decision-making (Batista, 2019). This is further supported by the expected utility theory, which offers a theoretical framework for understanding driver behavior in uncertain situations (Batista & Leclercq, 2019). This literature review synthesizes research from various disciplines to elucidate the mechanisms underlying traffic jam formation (Bayram, 2023). It examines five principal theoretical frameworks: heterogeneity in driver behavior, game-theoretic approaches, Nash equilibrium concepts, mathematical modeling techniques, and applications of expected utility theory (Chen, 2023). By integrating insights from these diverse approaches, this review aims to establish a foundation for further research on traffic flow optimization and intelligent transportation systems (Chen & Chen, 2023). The scope encompasses peer-reviewed research from journals on transportation engineering, operations research, behavioral economics, and applied mathematics (Cheng, 2019). A synthesis of themes and gaps in the current body of knowledge identifies promising directions for future research (Cheng & Yuan, 2019).

## 1.2. Driver behavior

Driver behavior fundamentally shapes traffic flow patterns and congestion (Dai and Filev, 2021). Uniformity in driver traits, preferences, and decision-making processes influences the overall performance of traffic systems. The complexity of human behavior in traffic settings encompasses risk tolerance, time preferences, social factors, and cognitive limitations (Fujino & Chen, 2019). Empirical research has identified specific driver archetypes that consistently exhibit similar behaviors across various traffic situations (Mahut, 2008). Aggressive drivers engage in potentially hazardous actions, such as tailgating, frequent lane changes, and accepting small gaps, which can lead to congestion cascades. However, cautious drivers, while individually safe, may cause system inefficiencies by reducing the throughput. Psychological studies on driver behavior have been conducted through both laboratory and field experiments (Massicot & Langbort, 2024). Factors such as stress, time pressure, and emotional states can influence driving decisions, sometimes resulting in suboptimal choices that lead to traffic jams (Takalloo, 2020). Social psychology research indicates that drivers often view other vehicles as competitors rather than collaborators in a shared transportation system. Cultural and demographic factors also play a role in shaping driver behavior patterns (Ye, 2017). Variables such as age, gender, driving experience, and culture affect risk perception, decision-making speed, and cooperation in traffic scenarios (Ye and Yang, 2017). These individual differences combine to create complex system-level dynamics that can either facilitate smooth traffic flow or contribute to congestion. Advances in data collection technologies now allow for a more sophisticated analysis of driver behavior in real-world settings. Tools such as GPRS tracking, smartphone sensors, and connected vehicle data provide unprecedented insights into actual driving patterns and decision-making processes (Zhu, 2019). These empirical findings challenge traditional assumptions about rational driver behavior and highlight the need for behavioral realism in traffic models.

## 1.3. Game theory

The application of game theory to traffic systems offers a mathematical framework for examining strategic interactions in which individual choices impact overall outcomes (Kan, 2019). When applied to transportation issues, game theory has uncovered essential insights into how traffic congestion emerges from drivers' strategic actions (Kan and Ferlis, 2019). Traffic scenarios inherently resemble strategic games, where the payoff is influenced by both the individual and other participants. The classic Prisoner's Dilemma has been utilized to understand traffic situations, such as lane selection and route choice (Lopetrone & Biondi, 2022).

Drivers often adopt strategies that maximize their personal benefits, even if these decisions lead to less optimal collective results (Wiersma, 1994). The tension between rationality and efficiency is central to many traffic congestion challenges (Wiersma and Stoop, 1994). Coordination games have been used to model situations in which drivers need to align their actions for smooth traffic flow (Ahn, 2007). Scenarios such as merging, managing intersections, and changing lanes require mutual cooperation for success (Ahn & Cassidy, 2007). These situations often face coordination challenges without direct communication among drivers (Brackstone, 1999). Evolutionary game theory illustrates how traffic behaviors evolve and persist through learning and adaptation. Drivers modify their strategies based on observed outcomes, leading to established traffic patterns and norms (Cassidy, 1999). This dynamic view explains why certain traffic behaviors become entrenched despite their inefficiency (Cassidy & Bertini, 1999). Mechanism design theory has been employed in traffic management strategies to align individual incentives with system-wide goals (Chen, 1991). Jankowski (1990) notes that "iterated playing by rational, self-interested actors will not result in cooperative behavior" (p.449). Mechanisms such as congestion pricing, dynamic tolling, and traffic signal optimization aim to influence driver behavior through incentives, relying on drivers' strategic responses to various incentive structures (Fagnant, 2015).

*1.4.* **Nash equilibrium**

The concept of Nash equilibrium is fundamental to comprehending stable traffic patterns that arise from strategic interactions among drivers (Fagnant & Kockelman, 2015). In traffic systems, Nash equilibrium occurs when no driver can improve their outcome by altering their strategy in response to the strategies of other drivers. This concept is particularly applicable to decisions regarding route selection, departure times, and lane choices (Garber and Gadiraju, 1989). Helbing (2001) extensively studied the existence and uniqueness of Nash equilibria in traffic scenarios. Research indicates that traffic systems often display multiple equilibria, each reflecting different driver behaviors and system performance patterns (Helbing, 1998). The presence of multiple equilibria complicates the prediction of traffic outcomes and the planning of management strategies (Helbing & Huberman, 1998). Wardrop's principles, formulated in the 1950s, represent one of the earliest applications of equilibrium concepts in traffic systems (Hoogendoorn, 2001). According to the user equilibrium principle, drivers select routes that minimize their individual travel costs, whereas the system optimum is the allocation that minimizes the total system cost, as defined by Hoogendoorn and Bovy (2001). The difference between the user equilibrium and system optimum outcomes quantifies the efficiency loss due to self-interested behavior, known as the price of anarchy. Nash equilibrium concepts have been dynamically extended to capture the temporal evolution of traffic patterns (Kesting, 2007). Dynamic user equilibrium models consider how drivers adjust their strategies over time in response to changing traffic conditions and experiences. These models reveal the stability and convergence properties of traffic systems (Kita, 1999). Computational methods have been developed for practical applications to identify Nash equilibria in large-scale traffic networks (Kockelman, 2005). These algorithms must manage the complexity of realistic traffic networks while considering strategic interactions among numerous drivers. Computational challenges have led to the development of approximation methods and heuristic approaches for equilibrium analysis.

1.5. **Mathematical and computational modeling**

Mathematical and computational modeling techniques offer a foundational analytical framework for understanding the formation of traffic jams and developing predictive capabilities (Laval & Daganzo, 2006). The evolution of traffic modeling has progressed from basic deterministic models to more complex, stochastic, and agent-based frameworks that simulate real-world traffic systems. Advances in computational power and the availability of large-scale traffic data have fueled these developments (Laval & Leclercq, 2013). Macroscopic traffic flow models treat traffic as a continuous fluid, emphasizing aggregate characteristics such as density, flow, and speed (Levinson, 2005). The fundamental relationship between flow and density, which underpins many macroscopic models, identifies the conditions that lead to traffic congestion (Li 2017). Models based on partial differential equations, such as the Lighthill-Whitham-Richards model, describe the wave-like propagation of traffic disturbances. Microscopic models, on the other hand, focus on the dynamics of individual vehicles and driver behavior, offering insights into the mechanisms behind traffic jams. The following models illustrate how drivers adjust their speed and spacing in response to the vehicle in front, while lane-changing models capture lateral movement decisions (Nagel & Schreckenberg, 1992). These microscopic models can replicate complex traffic phenomena, such as phantom jams and stop-and-go waves (Roughgarden 2005). In agent-based modeling approaches, drivers are treated as autonomous agents with unique characteristics and decision-making capabilities (Roughgarden 2002). These models can incorporate diverse driver behaviors, learning processes, and strategic interactions, which are often overlooked in conventional modeling frameworks (Roughgarden et al., 2002). Agent-based models are particularly effective for examining how individual behaviors contribute to collective traffic patterns. Stochastic modeling approaches embrace the inherent uncertainty of traffic systems by considering driver behavior and external conditions as random variables. Monte Carlo simulation methods enable the exploration of traffic system behavior under various scenarios and uncertain conditions (Stern, 2018). The use of machine learning and artificial intelligence techniques for pattern recognition and prediction is becoming increasingly prevalent in traffic modeling (Stern & Work, 2018).

**1.6. Expected utility theory**

Expected utility theory serves as the normative framework for comprehending how drivers make decisions in traffic situations under uncertainty. This theory posits that individuals opt for actions that maximize expected utility, where utility represents the subjective value of various outcomes (Sugiyama & Yukawa, 2008). In traffic scenarios, drivers decide on routes, departure times, and behaviors based on incomplete information about future traffic conditions. Applying the expected utility theory to traffic decision-making has provided new insights into driver behavior and system performance (Talebpour & Mahmassani, 2016). Typically, drivers tend to avoid risk by selecting routes with consistent travel times rather than those with greater variability. From a systemic perspective, this risk aversion can lead to suboptimal route choices, as drivers might bypass efficient but uncertain options. Prospect theory, as an alternative to expected utility theory, has been employed to explain systematic deviations from rational decision-making in traffic contexts (Treiber, 2013). This theory attributes drivers' preferences for familiar routes, even when superior alternatives exist, to loss aversion and reference dependence. These behavioral biases may account for persistent traffic patterns and resistance to traffic management strategies (Treiber, 2006). Multi-attribute utility theory has been used to model the trade-offs drivers make between different aspects of their travel experiences (Treiber & Helbing, 2006). When making travel decisions, drivers must weigh factors such as travel time, fuel costs, comfort, safety, and reliability (Wang, 2018). The significance of these attributes is subjective and may change depending on the trip's purpose and external conditions. By combining expected utility theory with game-theoretic models, a theoretical foundation has been established for examining strategic decision making in traffic systems (Wardrop, 1952). This integration allows for the analysis of how individual utility maximization leads to collective outcomes and the identification of conditions under which traffic congestion results from rational decision-making.

**1.7 Research Problem and Objectives**

This study investigates the primary research question: How do varying driver behavior archetypes contribute to the formation of traffic congestion? The objectives of this study include modeling the behaviors of passing and following using a substantial, randomly simulated population of drivers (n=500,000) with diverse behavior archetypes on a three-lane roadway, employing a game theory approach.

The research question is further explored in the context of how the expected utilities of drivers, and the total expected utilities of each traffic lane are affected when the proportion of each driver behavior archetype is modified. Specifically, scenarios were considered in which lanes consisted of 100% Responsible drivers, 100% Irresponsible drivers, and a 50% mix of Responsible and Irresponsible drivers. Utilizing the mathematical framework presented herein, a traffic jam model based on driver behavior was tested. The differences between the driver behavior archetypes were assessed using the Mann-Whitney U test.

**1.8 Significance and Contribution**

This research introduces an innovative approach to understanding traffic congestion from a game-theoretic perspective, emphasizing typical driver behaviors as identified by traffic violations. By integrating game-theoretic analysis with comprehensive simulations and robust nonparametric inference, this study reveals a significant imbalance in mixed driver populations: strategies of defection consistently surpass cooperative actions, leading to predictable congestion outcomes. This work reconceptualizes traffic flow as a strategic system in which the distribution of behaviors, rather than merely infrastructure, dictates performance, thereby extending beyond traditional models. For researchers, the framework offers testable predictions, scalable metrics (such as expected and final utilities), and a foundation for comparative analyses across archetypes, lanes, and policy settings. For practitioners, it provides practical diagnostics, such as forecasting congestion based on behavioral compositions, assessing interventions under stress, and formulating incentive-compatible policies that align individual motivations with collective efficiency. Collectively, these insights establish a robust foundation for behavior-aware traffic management and the development of evidence-based policies.

**2.1 Mathematical Framework**

The experiment is conducted within a specific framework, which is a variation of the chicken game from game theory (Jankowski, 1990). This framework posits that the lead vehicle in a lane typically encounters no obstructions, thereby allowing the driver to achieve full expected utility without any diminution. Drivers are presented with two options: either follow the vehicle ahead or overtake it, with each decision associated with a payoff matrix contingent upon their respective driver archetypes, resulting in a payoff and expected utility for each driver. If drivers choose to follow the vehicle ahead and share the same driver archetype, they may not experience any diminution in the expected utility. Conversely, if their archetypes differ, they might encounter a reduction in expected utility. This reduction can be construed as a penalty, as the following driver cannot fully realize the expected utility they would have in the absence of conflict with another driver. Consequently, the framework suggests that there will be an overall loss of expected utility in a lane if there are conflicting interests (to pass or follow) among different driver archetypes, with each driver endeavoring to optimally decide whether to follow or pass and maintain their original expected utility as if they were not delayed.

**2.2 Driver behavior archetypes**

Statistics on traffic violations have been used to estimate the prevalence of reckless driving behavior on roadways. Such actions are classified as aggressive driving by the National Highway Traffic Safety Administration (NHTSA) of the U.S. The Department of Transportation defines as "the operation of a motor vehicle in a manner that endangers or is likely to endanger persons or property" (Insurance Institute, 2025). Unfortunately, while the NHTSA records fatal accidents, comprehensive data on non-fatal traffic offenses are not readily accessible. However, conviction data for various aggressive or hazardous traffic violations from Ontario, Canada (2023) facilitate the categorization of these behaviors. In Table 1, the most prevalent offense was failing to share the road, accounting for 29.36% of the selected violations, followed by improper use of the high-occupancy vehicle (HOV) lane, which was typically located on the far left side of the highway (22.91 %), following too closely (23.34 %), and careless driving (17.43 %). Although these behaviors can result in accidents involving injuries, fatalities, and property damage, they also pose a risk and inconvenience to other road users who adhere to traffic laws and drive responsibly, even in the absence of an accident. More importantly, these actions affect traffic flow and the overall driving experience, whether traffic is smooth or disrupted by irresponsible behavior.

**Table 1. Conviction data – traffic offenses, Ontario, Canada.**

| Traffic offense | # of convictions | % of behavior total |
|---|---|---|
| Careless driving | 1,358 | 17.43% |
| Fail to share road-passing, meeting others | 144 | 1.85% |
| Fail to share road-when overtaken | 8 | 0.10% |
| Fail to share road-when overtaking | 2280 | 29.26% |
| Following too closely-motor vehicle | 1819 | 23.34% |
| Following too closely. | 7 | 0.09% |
| Improper passing | 8 | 0.10% |
| Improper passing-overtaking traffic | 147 | 1.89% |
| Improper use of high occupancy vehicle lane (HOV) | 1785 | 22.91% |
| Pass on right-unsafe conditions/off roadway | 129 | 1.66% |
| Unnecessary slow driving | 108 | 1.39% |
| Offense total | 7,793 | 100.00% |

**Source: Ontario (2023).**

The framework delineates drivers into two principal categories based on their behavior: Responsible and Irresponsible drivers. In our model, responsible drivers are defined by their compliance with traffic regulations, avoidance of erratic driving, and exhibition of courtesy and caution when overtaking or following other vehicles. In contrast, irresponsible drivers are those who disregard legal regulations, engage in hazardous driving behaviors, and execute maneuvers that pose a risk to themselves or others, often driving with self-serving motives or to the detriment of fellow drivers. Within the Irresponsible drivers category, five key archetypes of irresponsible behavior are identified using traffic offense data. Table 2 summarizes the driver behavior archetypes.

**Table 2. The driver behavior archetypes are based on irresponsible driver behavior.**

| Driver behavior archetype | Behaviors corresponding to traffic offenses | % |
|---|---|---|
| Impatient drivers | Following too closely, improper passing, pass on right unsafe conditions | 27.08% |
| Left lane camper | Improper use of HOV lane | 22.91% |
| Slowpokes | Unnecessary slow driving | 1.39% |
| Selfish drivers | Fail to share road | 31.21% |
| Dangerous drivers | Careless driving | 17.43% |
| Total | | 100.00 |

**Source: summary of Table 1**

The driver archetypes were randomly generated and coded according to the scheme presented in Table 3. With the exception of the Left Lane Campers, which consistently occupy the left lane (Lane 1), all other driver archetypes were randomly assigned to any of the three lanes.

**Table 3. Driver behavior archetype coding and lane assignment for the simulation.**

| τ | Driver Archetype | Lane assignment |
|---|---|---|
| 0 | Responsible | Occupy any lane |
| 1 | Selfish | Occupy any lane |
| 2 | Impatient | Occupy any lane |
| 3 | Left Lane Camper | Always occupy left lane |
| 4 | Dangerous | Occupy any lane |
| 5 | Slowpoke | Occupy any lane |

**Source: simulation assumption**

## 2.3 The Driver Behavior Archetype Model

The driver behavior archetype model underpins a three-lane traffic simulation designed to examine the impact of driver behavior composition on the overall utility and traffic flow dynamics. This model formalizes strategic interactions among diverse driver types and delineates the mathematical conditions under which traffic congestion emerges from individual driver decisions. The simulation classifies drivers into two primary categories based on their behavior and adherence to rules. Responsible drivers (denoted $\mathbb{R}$) represent the cooperative driver model, consistently adhering to speed limits, maintaining safe distances, changing lanes only when safe, and respecting the rights of other drivers. This driver type is associated with a baseline utility of u_base($\mathbb{R}$) = 5, indicating the enhanced individual and collective benefits of cooperative behavior. In contrast, irresponsible drivers (denoted $\mathcal{J}$) encompass five distinct subtypes characterized by aggressive or non-compliant behavior: selfish drivers who prioritize personal gain, dangerous drivers who take excessive risks, left-lane campers who obstruct faster traffic, slowpokes who impede flow, and impatient drivers who make rash decisions. These irresponsible subtypes share traits such as violating speed limits, executing unsafe passing maneuvers, and disregarding traffic rules, with a baseline utility of u_base($\mathcal{J}$) = 3, reflecting the reduced long-term benefits of non-cooperative strategies.

In the simulation, for each driver i positioned at i > 1 (subsequent drivers), the expected utility $u_i^E$ was determined by considering the probabilistic actions of both the leading driver (i-1) and the following driver (i). The formula for expected utility, $u_i^E = \Sigma_{a,b} p^{(\tau_{i-1},\tau_i)}(a) \cdot p^{(\tau_{i-1},\tau_i)}(b) \cdot M^{(\tau_{i-1},\tau_i)}[a,b]$, sums over all potential action pairs (a,b), with each pair weighted by the likelihood of each driver type choosing each action. These probabilities are based on the behavioral traits of each driver type, where responsible drivers are more likely to engage in cooperative actions, while irresponsible drivers tend to choose aggressive actions. The lead driver in the first position receives only the baseline utility $u_1^E = u_{base}(\tau_1)$ because they do not have any following constraints, highlighting the benefit of being first in the lane.

A fundamental component of the simulation is the adjustment penalty applied when consecutive drivers are of different types. The adjusted utility $u_i^A$ is equivalent to the expected utility $u_i^E$ when the leading and following drivers are of the same type ($\tau_{i-1} = \tau_i$), reflecting the compatibility and predictability of interactions among similar drivers. Conversely, when driver types differ ($\tau_{i-1} \neq \tau_i$), a penalty of 0.5 units is imposed: $u_i^A = u_i^E - 0.5$. This penalty accounts for coordination costs, increased uncertainty, and potential safety risks associated with interactions between drivers exhibiting fundamentally different behaviors. In scenarios with mixed driver types, this penalty is frequently applied as Responsible and Irresponsible drivers alternate in the lane, leading to a systematic reduction in utility. In scenarios where all drivers are either Responsible or Irresponsible, this penalty is never applied, resulting in more stable (though not necessarily higher) utility trajectories.

The model delineates a critical threshold condition that determines when a traffic system transitions from a flowing to a jammed state. The simulation calculates the total adjusted utility ($U_{\text{total}} = \Sigma_{i=1}^n u_i^A$) for all $n$ drivers in the lane. If $U_{\text{total}}$ is less than or equal to zero, the traffic jam condition is triggered, capping all driver utilities at zero or below: $u_i^J = \min\{u_i^A, 0\}$ for each driver $i$. This condition signifies the point at which the cumulative negative effects of driver interactions surpass the capacity of the system to maintain flow. Once this threshold is reached, no driver can achieve positive utility, irrespective of their individual actions, thereby formalizing traffic breakdown as a system-level emergent phenomenon. The simulation monitors this condition separately for each of the three lanes, illustrating how different lane positions and driver compositions influence the threshold at which a breakdown occurs.

The final utility for each driver is contingent on the traffic jam condition $u_i^{final} = u_i^J$ if the jam condition is satisfied; otherwise, $u_i^{final} = u_i^A$. The total lane utility, $U_\mathscr{L} = \sum_{i=1}^n u_i^{final}$, encapsulates the collective welfare of all drivers within the lane and serves as the principal outcome measure in the simulation. The findings indicate that $U_\mathscr{L}$ exhibits significant variation across the three scenarios. In the all-Responsible scenario, the absence of type mismatch penalties and elevated baseline utilities culminates in the highest total lane utility and the most stable utility trajectories. Conversely, in the mixed scenario, frequent-type mismatches result in systematic penalties, leading to a more rapid degradation of utility as additional drivers enter the system. In the all-irresponsible scenario, lower baseline utilities and aggressive interaction patterns precipitate the steepest utility decline and the earliest onset of traffic jam conditions, notwithstanding temporarily higher throughput in Lane 1.

## 2.4 Driver behavior archetype model - formalized

The driver behavior archetype model is formally presented and defined as follows:

**Definition 1.1 (Driver classification and lane definition)**

Let $\mathscr{L} = \{d_1, d_2, \ldots, d_n\}$ be a lane with n drivers in positions 1 to n.

Let $\tau_i \in \{\mathscr{R}, \mathscr{J}\}$ be the type of driver i, where $\mathscr{R}$ denotes Responsible and $\mathscr{J}$ denotes Irresponsible. All Responsible drivers obey the speed limit, pass and drive safely, and obey the rules of the road, respecting all the other drivers. Irresponsible drivers consist of a set of aggressive or irresponsible driver types: Selfish, Dangerous, Left Lane Camper, Slowpoke, and Impatient. These drivers are members of the set $\mathscr{J}$. These drivers do not obey the speed limit, pass, and driver in an unsafe manner, and do not obey the rules of the road. Irresponsible drivers ($\mathscr{J}$) encompass a set of aggressive or problematic driving behaviors, including selfish drivers who prioritize personal advancement, dangerous drivers who engage in risky maneuvers, left-lane campers who impede traffic flow, slowpokes who drive significantly below optimal speeds, and impatient drivers who make aggressive lane changes.

Let $M^{(\alpha,\beta)}$ be the payoff matrix for the interaction between the lead driver of type α and the following driver of type β.

Let $p^{(\alpha,\beta)}(a)$ be the probability that a following driver of type β performs an action when following a lead driver of type α.

Let $u\_base(\tau)$ be the baseline utility for driver type τ.

**Definition 1.2 (Payoff Matrices)**

The payoff matrices are defined as

$$M^{(\mathscr{R},\mathscr{R})} = \begin{bmatrix} 0 & -1 \\ 1 & -1 \end{bmatrix} \text{ (Responsible vs Responsible)}$$

$$M^{(\mathscr{R},\mathscr{J})} = \begin{bmatrix} 1 & 1 \\ -1 & -1 \end{bmatrix} \text{ (Responsible vs Irresponsible)}$$

$$M^{(\mathscr{J},\mathscr{J})} = \begin{bmatrix} 0 & 1 \\ 1 & -1 \end{bmatrix} \text{ (Irresponsible vs Irresponsible)}$$

**Source: Payoff matrices adapted from Stanford (1998).**

where rows correspond to lead driver actions and columns to following driver actions.

**Definition 1.3 (Baseline Utilities)**

$u\_base(\mathscr{R}) = 5, \quad u\_base(\mathscr{J}) = 3$

**Lemma 1.1 (Lead Driver Utility)**

For the lead driver $d_1$:

Expected utility: $u_1^E = u\_base(\tau_1)$

Adjusted utility: $u_1^A = 0$ (not following anyone, no blocking effects)

### Lemma 1.2 (Following Driver Expected Utility)

For following driver $d_i$ where $i > 1$:

$$u_i^E = \sum_{a,\beta} p^\wedge(\tau_{i-1},\tau_i)(a) \cdot p^\wedge(\tau_{i-1},\tau_i)(b) \cdot M^\wedge(\tau_{i-1},\tau_i)[a,b]$$

where the summation is over all possible action pairs (a,b).

The probability that a lead driver of type α chooses action a and a following driver of type β chooses action b is given by the probability matrix for these actions in Table 4. These probabilities were assumed.

**Table 4. Probability of driver action: assumptions.**

| Lead driver | Following driver | p(follow) | p(pass) |
|---|---|---|---|
| Responsible | Responsible | 0.5 | 0.5 |
| Responsible | Irresponsible | 0.1 | 0.9 |
| Irresponsible | Responsible | 0.3 | 0.7 |
| Irresponsible | Irresponsible | 0.1 | 0.9 |

Source: model assumptions

### Lemma 1.3 (Adjustment Rules)

For following driver $d_i$ where $i > 1$:

$$u_i^A = \begin{cases} u_i^E & \text{if } \tau_{i-1} = \tau_i \text{ (same type, no penalty)} \\ u_i^E - 0.5 & \text{if } \tau_{i-1} \neq \tau_i \text{ (different types, penalty applied)} \end{cases}$$

### Lemma 1.4 (Traffic Jam Condition)

Let $U\_total = \sum_{i=1}^{n} u_i^A$ be the sum of all adjusted utilities.

If $U\_total \leq 0$, then the traffic jam condition applies:

$u_i^J = \min\{u_i^A, 0\} \quad \forall i \in \{1, 2, \ldots, n\}$

### Theorem 1.1 (Driver Utility Analysis)

*The total utility of lane $\mathscr{L}$ is given by*

$U\_\mathscr{L} = \Sigma_{i=1}^{n} u_i^{final}$

*where*

$u_i^{final} = \begin{cases} u_i^J & \text{if traffic jam condition applies} \\ u_i^A & \text{otherwise} \end{cases}$

*More explicitly:*

$u_i^{final} = \begin{cases} \min\{u_i^A, 0\} & \text{if } \Sigma_{j=1}^{n} u_j^A \leq 0 \\ u_i^A & \text{if } \Sigma_{j=1}^{n} u_j^A > 0 \end{cases}$

**Proof Outline**

The theorem follows directly from the composition of Lemmas 1.1-1.4:

1. By Lemma 1.1, the lead driver utility is determined solely by the baseline type.

2. By Lemma 1.2, the following driver utilities depend on strategic interactions via payoff matrices.

3. By Lemma 1.3, type mismatches incur adjustment penalties.

4. By Lemma 1.4, system-wide negative utility triggers traffic jam conditions.

The total utility $U\_\mathscr{L}$ is thus the sum of individual final utilities, where each $u_i^{final}$ is determined by the conditional structure defined above.

**Corollary 1.1**

*Under the assumptions of Theorem 1.1, lane exhibits the following properties.*

*(i)* $U\_\mathscr{L} \leq \Sigma_{i=1}^{n} u\_base(\tau_i)$ *(total utility bounded above by baseline sum)*

*(ii) If all drivers are of the same type, then $u_i^A = u_i^E$ for all $i > 1$*

*(iii) Type heterogeneity introduces systematic utility penalties of magnitude 0.5 per mismatch*

**Corollary 1.2**

*Degradation patterns that correlate with traffic congestion emergence occur when the total expected utility of lane $\mathscr{L}$ meets the condition*

$$\Sigma_{j=1} EU(d_i) \leq 0.$$

## 3. Materials and Methods

This study employed data generated via Julius.ai, a platform for data analysis that utilizes artificial intelligence. In three distinct experiments, 500,000 driver types were randomly generated based on six behavioral archetypes. These archetypes were formulated by analyzing traffic offense conviction data and categorizing offense types into broader behavioral groups. The sample size was selected to represent the average daily traffic volume on busy highways. Each driver was randomly assigned to one of three lanes: left lane (lane 1), middle lane (lane 2), or right lane (lane 3). In the first experiment, the driver types were evenly divided between 50% irresponsible and 50% responsible drivers. The second experiment comprised entirely 100% responsible drivers, whereas the third experiment involved 100% irresponsible drivers. Payoff matrices and expected utilities for each driver were calculated and adjusted according to their positions, whether they were leading or following a driver of a specific behavioral archetype. The cumulative expected utility and adjusted expected utility for each lane were documented. Figure 1 illustrates the simulation framework.

**Figure 1.** Simulation framework.

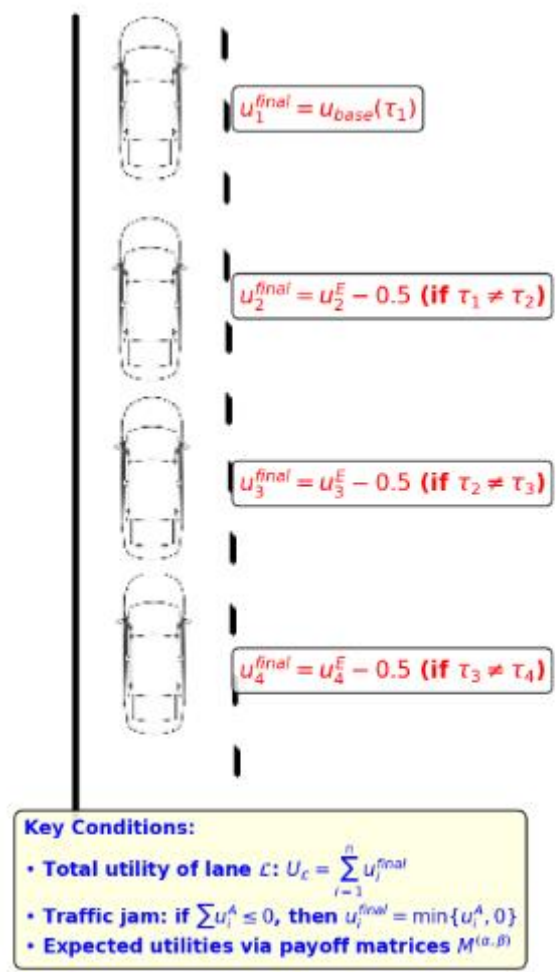

**Source: model illustration**

## 4. Results

The sample data were analyzed.

### *4.1. Descriptive statistics*

Table 5 presents the distribution of drivers in a scenario in which the sample consists of 50% irresponsible and 50% responsible drivers. Within the sample of 500,000 drivers, Responsible drivers constituted exactly half of the population (250,000; 50.0%), with an almost equal distribution across lanes (Lane 1:83,560, 16.7%; Lane 2:83,121, 16.6%; Lane 3:83,319, 16.7%). Among the irresponsible archetypes, Selfish (78,022; 15.6%) and Impatient (67,688; 13.5%) drivers were the most prevalent, each demonstrating balanced representation across lanes. Dangerous drivers accounted for 8.7% of the total drivers (43,564), again evenly distributed by lane. Slowpokes are infrequent (0.7% overall), with minimal variation across lanes. A notable structural asymmetry was observed in the concentration of Left Lane Campers exclusively in Lane 1 (57,262; 11.5%), with none in Lanes 2 or 3, indicating a lane-specific behavioral niche consistent with the role of the left lane in passing and queuing dynamics. Aggregate lane totals revealed a higher utilization of Lane 1 (204,531; 40.9%) compared to the middle and right lanes (29.5% and 29.6%, respectively), suggesting that the left lane attracts a disproportionately large share of both Responsible drivers and lane-specific behaviors (e.g., Left Lane Camping). Overall, the distribution exhibits broad cross-lane symmetry for most archetypes, with the singular exception of left-lane campers, whose concentration in the left lane may have significant implications for the flow, overtaking, and expected-utility dynamics modeled by the model.

**Table 5. Distribution of driver types by lane. 50% Responsible and 50% Irresponsible drivers.**

| Driver Type | Lane 1 (Left) | Lane 2 (Middle) | Lane 3 (Right) | Total (All Lanes) |
|---|---|---|---|---|
| Dangerous | 14,352 (2.9%) | 14,589 (2.9%) | 14,623 (2.9%) | 43,564 (8.7%) |
| Impatient | 22,469 (4.5%) | 22,574 (4.5%) | 22,645 (4.5%) | 67,688 (13.5%) |
| Left Lane Camper | 57,262 (11.5%) | 0 (0.0%) | 0 (0.0%) | 57,262 (11.5%) |
| Responsible | 83,560 (16.7%) | 83,121 (16.6%) | 83,319 (16.7%) | 250,000 (50.0%) |
| Selfish | 25,770 (5.2%) | 26,181 (5.2%) | 26,071 (5.2%) | 78,022 (15.6%) |
| Slowpoke | 1,118 (0.2%) | 1,176 (0.2%) | 1,170 (0.2%) | 3,464 (0.7%) |
| Total | 204,531 (40.9%) | 147,641 (29.5%) | 147,828 (29.6%) | 500,000 (100.0%) |

**Source: data analysis**

### 4.2. Expected utility by driver type

In Table 6, the mean final expected utility is analyzed across different driver type categories and lanes. Irresponsible drivers ($\mathscr{J}$) demonstrated a higher final expected utility compared to Responsible drivers ($\mathscr{R}$) in each lane and overall. Specifically, for $\mathscr{J}$, the mean final expected utility is -0.204 in Lane 1, −0.282 in Lane 2, and −0.282 in Lane 3, resulting in an overall mean of −0.244. In contrast, for $\mathscr{R}$, the corresponding values were −0.398 (Lane 1), −0.360 (Lane 2), and −0.359 (Lane 3), with an overall mean of −0.372. The lane pattern suggests a relative advantage for lane 1 in both categories (less negative utility), with lanes 2 and 3 closely aligned and more negative than lane 1. Notably, the disparity between categories is most pronounced in Lane 1 ($\Delta \approx 0.193$) and smaller yet consistent in Lanes 2 and 3 ($\Delta \approx 0.077$–$0.080$), culminating in a significant overall difference favoring $\mathscr{J}$ (overall $\Delta \approx 0.128$). Sample sizes were balanced across categories (n = 250,000 per category), with a greater number of observations in Lane 1 than in Lanes 2 and 3 for both $\mathscr{J}$ (Lane 1: 120,971; Lane 2: 64,520; Lane 3: 64,509) and $\mathscr{R}$ (Lane 1: 83,560; Lane 2: 83,121; Lane 3: 83,319).

Collectively, these findings suggest that in a mixed environment, irresponsible behavior systematically achieves a higher final expected utility than responsible behavior, particularly in Lane 1, while the middle and right lanes exhibit nearly identical performance profiles within each category.

**Table 6. Mean final expected utility by driver type category and lane. 50% Responsible and 50% Irresponsible drivers.**

| Driver type category | Lane 1 | Lane 2 | Lane 3 | Overall | Lane 1 count | Lane 2 count | Lane 3 count | Overall count |
|---|---|---|---|---|---|---|---|---|
| Irresponsible ($\mathscr{I}$) | -0.2041 | -0.2824 | -0.2821 | -0.2444 | 120971 | 64520 | 64509 | 250000 |
| Responsible ($\mathscr{R}$) | -0.3978 | -0.3596 | -0.3592 | -0.3722 | 83560 | 83121 | 83319 | 250000 |

**Source: data analysis**

### 4.3. Traffic congestion emergence

Figure 2 illustrates the cumulative expected utilities for the initial 25 drivers in each lane for the 50% Responsible and 50% Irresponsible scenarios. According to this model, a traffic jam commences when the total expected utility of a lane reaches zero or falls below zero. Initially, Lane 1 (the left lane) maintains a positive cumulative expected utility for a brief period; however, it declines to zero and becomes increasingly negative by the time it reaches driver number 10. Lane 3 ( right lane) exhibited the poorest performance. The cumulative expected utility for Lane 2 (the middle lane) aligns with that of Lane 3 around Driver 7. Notably, the cumulative expected utilities for both Lanes 2 and 3 remain quite similar up to approximately driver number 15, after which Lane 2's expected utility surpasses that of Lane 3, despite decreasing at a slightly slower rate than Lane 3. This suggests that drivers closest to the lead car in a lane experience better expected utilities than those farther back in the line, particularly in Lane 1.

**Figure 2. Expected utility running total by driver position, first 25 cars by lane, 50% Responsible versus 50% Irresponsible drivers.**

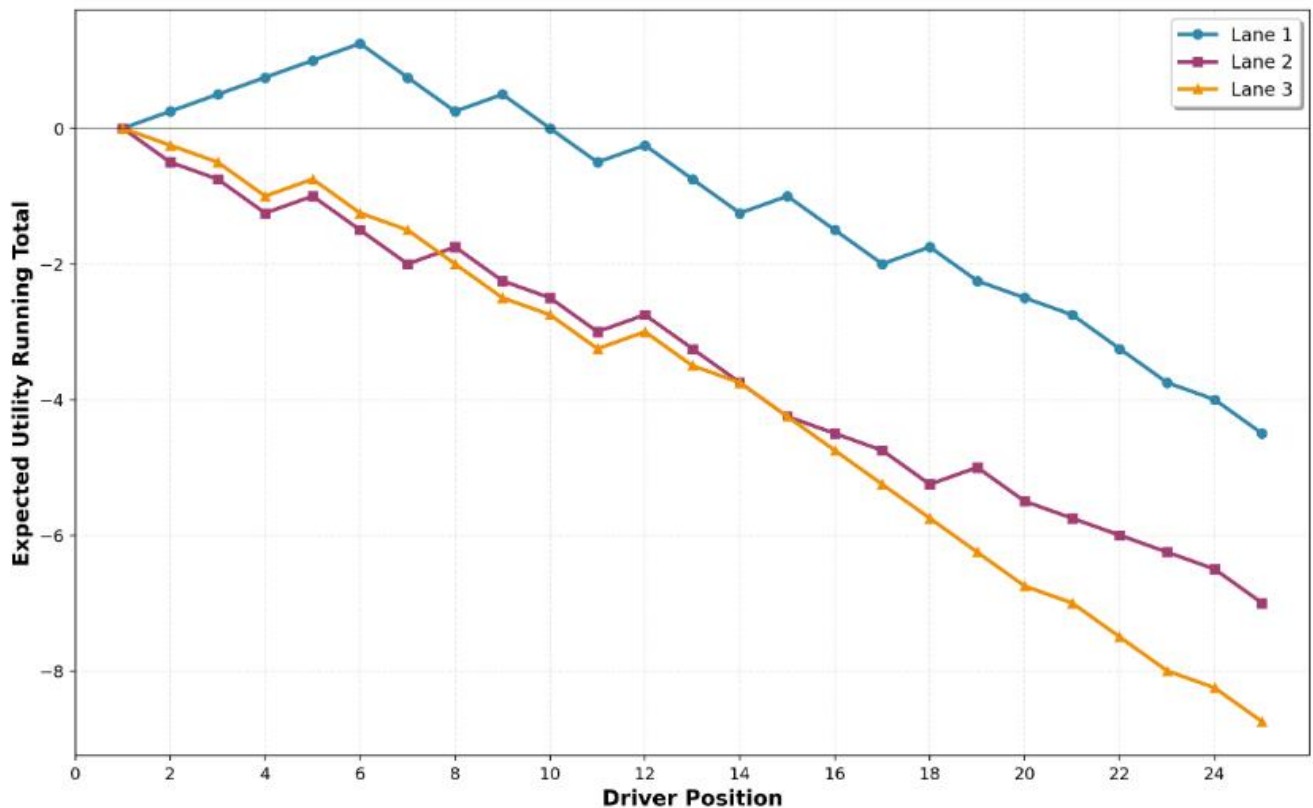

**Source: data analysis**

The Mann-Whitney test has been widely employed in research on traffic behavior (Gurupakiam and Jones, 2012; Zulkifli and Ponrahono, 2018; Li and Cheng, 2019). This statistical method was utilized to evaluate the differences in expected utilities, initial or original payoffs, and final utilities between Irresponsible ($\mathcal{J}$) and Responsible drivers ($\mathcal{R}$). The null hypothesis posits that there is no difference in the mean expected utilities, initial or original payoffs, and final utilities between Irresponsible ($\mathcal{J}$) and Responsible drivers ($\mathcal{R}$), while the alternative hypothesis suggests that a difference exists in these values for Irresponsible ($\mathcal{J}$) and Responsible drivers ($\mathcal{R}$). The results presented in Table ? demonstrate significant distinctions between Responsible and Irresponsible drivers in terms of their mean expected utilities, original payoffs, and final utilities. The effect sizes between the group means were assessed and interpreted using Cohen's d statistic (Cohen, 1988) and the effect size r statistic (Cohen, 1988; Rosnow and In the simulation, for each driver i positioned at i > 1 (subsequent drivers), the expected utility $u_i^E$ was determined by considering the probabilistic actions of both the leading driver (i-1) and the following driver (i).Let ^al, 2003).The assumptions necessary for the application of the Mann–Whitney U test are satisfied, as outlined by Fay and Proschan (2010). Independence is ensured through random sampling of drivers who serve as independent observational units. The measurements were conducted on a continuous ratio scale, thereby fulfilling the ordinal/continuous requirement. The sample sizes are substantial, with N = 250,000 per group, surpassing the typical thresholds required for asymptotic procedures (Conover 1999; Gibbons and Chakraborti 2011). Ties are addressed using SciPy's implementation, which employs suitable exact or asymptotic methods and tie corrections, as detailed in the API reference (SciPy, 2025; Virtanen et al., 2020), aligned with standard nonparametric methodologies (Conover, 1999). Furthermore, the group distributions exhibit similar shapes, which supports the valid interpretation of the Mann–Whitney U test as a comparison of the central tendency under matched distributional forms (Gibbons & Chakraborti, 2011).

Table 7 presents a series of Mann-Whitney U tests to compare responsible and irresponsible driver populations in a mixed traffic environment (50% responsible, 50% irresponsible) across three key variables: Expected Utility, Original Payoff, and Final Utility. All comparisons revealed statistically significant differences (p <.001). For Expected Utility, irresponsible drivers (M = 0.128) demonstrated higher values than responsible drivers (M = -0.127), with a mean difference of -0.256, U = 7.47 × 10$^9$, p < .001, d = -2.039. This large negative effect size indicates that in mixed traffic conditions, irresponsible drivers paradoxically achieve higher expected utility, suggesting that they exploit the cooperative behavior of responsible drivers. The original payoff showed that responsible drivers (M = 2.999) significantly outperformed irresponsible drivers (M = -0.007), with a mean difference of 3.007, U = 6.25 × 10$^{10}$, p < .001, d = 51.815, representing a massive effect. This demonstrates that fundamental game-theoretic payoffs strongly favor responsible behavior. However, Final Utility revealed a more nuanced pattern, with irresponsible drivers (M = -0.244) achieving slightly higher outcomes than responsible drivers (M = -0.372), yielding a mean difference of -0.128, U = 2.31 × 10$^{10}$, p < .001, d = -0.647. This medium effect size (r = 0.226) suggests that while responsible drivers receive superior original payoffs, the final utility calculations partially offset this advantage when irresponsible drivers are present in the system. The contrasting effect sizes across variables, ranging from medium (d = -0.647) to massive (d = 51.815), highlight the complex dynamics of mixed-traffic populations. These findings suggest that while responsible driving behavior yields superior original payoffs, irresponsible drivers may exploit cooperative environments to achieve competitive or even superior expected and final utilities, representing a classic social dilemma in which individual rationality conflicts with collective welfare.

**Table 7. Mann-Whitney test results: 50% irresponsible ($\mathscr{J}$) versus 50% Responsible ($\mathscr{R}$) drivers.**

| Variable | $\mathscr{R}$ Mean | $\mathscr{J}$ Mean | Mean Difference | U Statistic | P Value | Cohen's d | Effect Size r | Effect Category |
|---|---|---|---|---|---|---|---|---|
| Expected Utility | -0.127 | 0.128 | -0.256 | 7.47E+09 | < 0.001 | -2.039 | 0.659 | Large |
| Original Payoff | 2.999 | -0.007 | 3.007 | 6.25E+10 | < 0.001 | 51.815 | 0.866 | Large |
| Final Utility | -0.372 | -0.244 | -0.128 | 2.31E+10 | < 0.001 | -0.647 | 0.226 | Medium |

**Source: data analysis**

Table 8 presents the results of a series of Mann-Whitney U tests conducted to compare the 100% Responsible and 100% Irresponsible driver populations across three key variables: Model Expected Utility, Original Payoff, and Final Utility. All comparisons revealed statistically significant differences (p <.001). For Model Expected Utility, responsible drivers (M = -0.090) demonstrated substantially higher values than irresponsible drivers (M = -1.470), with a mean difference of 1.380, U = 2.50 × $10^{11}$, p < .001, d = 28.55. This extremely large effect size indicates that model-adjusted utility strongly favors responsible driving behavior. Similarly, Original Payoff showed responsible drivers (M = 0.000) significantly outperforming irresponsible drivers (M = -1.000), with a mean difference of 1.000, U = 2.50 × $10^{11}$, p < .001, d = 258.20, representing a massive effect. Finally, Final Utility maintained this pattern, with responsible drivers (M = -0.090) achieving higher outcomes than irresponsible drivers (M = -1.470), yielding a mean difference of 1.380, U = 2.50 × $10^{11}$, p < .001, d = 29.39. All three variables exhibited Cohen's d values far exceeding the conventional threshold of 0.8 for large effects, with U statistics approaching 2.5 × $10^{11}$ indicating nearly complete separation between the two distributions.

**Table 8. Mann-Whitney test results: 100% irresponsible ($\mathscr{I}$) versus 100% Responsible ($\mathscr{R}$) drivers.**

| Variable | $\mathscr{R}$ Mean | $\mathscr{I}$ Mean | Mean Difference | U Statistic | P Value | Cohen's d | Effect Size r | Effect Category |
|---|---|---|---|---|---|---|---|---|
| Expected Utility | -0.090 | -1.470 | 1.380 | 2.5E+11 | < 0.001 | 28.555 | 1.000 | Large |
| Original Payoff | 0.000 | -1.000 | 1.000 | 2.5E+11 | < 0.001 | 258.198 | 1.000 | Large |
| Final Utility | -0.090 | -1.470 | 1.380 | 2.5E+11 | < 0.001 | 29.392 | 1.000 | Large |

**Source: data analysis**

Table 9 presents a series of Mann-Whitney U test results to investigate pairwise differences in expected utility across six driver archetypes within a mixed traffic environment comprising 50% responsible and 50% irresponsible drivers. The analysis revealed a distinct bifurcation in outcomes: comparisons involving Responsible drivers versus any irresponsible archetype yielded statistically significant differences with large effect sizes, whereas comparisons among irresponsible archetypes showed no meaningful differences. Specifically, Dangerous drivers (M = 0.149) significantly outperformed responsible drivers (M = -0.102), with a mean difference of 0.251, U = 1.06 × 10⁻, p < .001, d = 2.043, r = 0.880. Similarly, Impatient drivers (M = 0.147) demonstrated significantly higher expected utility than Responsible drivers (M = -0.102), yielding a mean difference of 0.250, U = 1.65 × 10⁹, p < .001, d = 2.029, r = 0.879. Left-lane campers (M = 0.148) also significantly exceeded responsible drivers (M = -0.102) in expected utility, with a mean difference of 0.250, U = 4.21 × 10⁹, p < .001, d = 2.031, r = 0.879. Conversely, when Responsible drivers were compared to Selfish drivers (M = 0.148), the pattern reversed, showing Responsible drivers with significantly lower expected utility, mean difference = -0.250, U = 2.59 × 10⁸, p < .001, d = -2.039, r = 0.120, and similarly for Slowpoke drivers (M = 0.148), mean difference = -0.251, U = 1.12 × 10⁷, p < .001, d = -2.041, r = 0.120. All comparisons among irresponsible archetypes (Dangerous vs. Impatient, Dangerous vs. Left Lane Camper, Dangerous vs. Selfish, Dangerous vs. Slowpoke, Impatient vs. Left Lane Camper, Impatient vs. Selfish, Impatient vs. Slowpoke, Left Lane Camper vs. Selfish, Left Lane Camper vs. Slowpoke, and Selfish vs. Slowpoke) revealed no statistically significant differences, with p values ranging from .287 to .956, Cohen's d values between -0.010 and 0.011, and effect sizes r clustered tightly around 0.500, indicating complete distributional overlap. These findings demonstrate that in mixed traffic conditions, all irresponsible driver archetypes achieve statistically equivalent expected utilities (M ≈ 0.148), while responsible drivers experience systematically lower outcomes (M = -0.102). The consistently large effect sizes (d ≈ 2.0, r ≈ 0.88) for all Responsible versus irresponsible comparisons, coupled with negligible effects among irresponsible types, suggest that the model captures a fundamental asymmetry: irresponsible behaviors exploit cooperative environments uniformly, regardless of specific tactical differences (e.g., aggression, lane discipline, speed), whereas responsible behavior incurs systematic costs when embedded in mixed populations. This pattern supports the interpretation of mixed traffic as a social dilemma in which defection strategies converge to a common payoff advantage over cooperation, irrespective of the particular form of defection employed.

**Table 9. Mann-Whitney test results – driver archetype pairings – 50% irresponsible versus 50% responsible drivers.**

| Driver Type 1 | Driver Type 2 | Driver Type 1 Mean | Driver Type 2 Mean | Mean Difference | U Statistic | P Value | Cohen's d | Effect Size r |
|---|---|---|---|---|---|---|---|---|
| Dangerous | Impatient | 0.149 | 0.147 | 0.001 | 162139348 | 0.287 | 0.011 | 0.503 |
| Dangerous | Left Lane Camper | 0.149 | 0.148 | 0.001 | 412272468.0 | 0.471 | 0.006 | 0.502 |
| Dangerous | Responsible | 0.149 | -0.102 | 0.251 | 1055638632.0 | 0.000 | 2.043 | 0.880 |
| Dangerous | Selfish | 0.149 | 0.148 | 0.000 | 185255424 | 0.727 | 0.004 | 0.501 |
| Dangerous | Slowpoke | 0.149 | 0.148 | 0.000 | 8029528.0 | 0.956 | 0.002 | 0.500 |
| Impatient | Left Lane Camper | 0.147 | 0.148 | -0.001 | 641841686.3 | 0.555 | -0.005 | 0.499 |
| Impatient | Responsible | 0.147 | -0.102 | 0.250 | 1649568557.0 | 0.000 | 2.029 | 0.879 |
| Impatient | Selfish | 0.147 | 0.148 | -0.001 | 288410300.5 | 0.396 | -0.008 | 0.498 |
| Impatient | Slowpoke | 0.147 | 0.148 | -0.001 | 12500505.0 | 0.753 | -0.010 | 0.498 |
| Left Lane Camper | Responsible | 0.148 | -0.102 | 0.250 | 4207119985.0 | 0.000 | 2.031 | 0.879 |
| Left Lane Camper | Selfish | 0.148 | 0.148 | 0.000 | 736694494.5 | 0.679 | -0.003 | 0.499 |
| Left Lane Camper | Slowpoke | 0.148 | 0.148 | -0.001 | 31930459.0 | 0.868 | -0.005 | 0.499 |
| Responsible | Selfish | -0.102 | 0.148 | -0.250 | 259004877.0 | 0.000 | -2.039 | 0.120 |
| Responsible | Slowpoke | -0.102 | 0.148 | -0.251 | 11210622.0 | 0.000 | -2.041 | 0.120 |
| Selfish | Slowpoke | 0.148 | 0.148 | 0.000 | 14391869.0 | 0.950 | -0.002 | 0.500 |

**Source: data analysis**

Figures 3, 4, and 5 illustrate the cumulative expected utilities for each lane under three distinct driver behavior scenarios: 100% Responsible, 50% Responsible-50% Irresponsible, and 100% Irresponsible, respectively. The figures consistently reveal that the 100% Responsible scenario exhibits the smallest decline in the total expected utilities as the number of drivers increases, whereas the 100% Irresponsible scenario exhibits the most pronounced slope. Notably, the slope of the cumulative expected utility line for the 50% Responsible-50% Irresponsible scenario in lane 1 (left lane) closely resembled that of the 100% Responsible scenario. However, the slopes for the 50% Responsible-50% Irresponsible scenario in lanes 2 and 3 were steeper.

**Figure 3. Running total of expected driver utility for lane 1.**

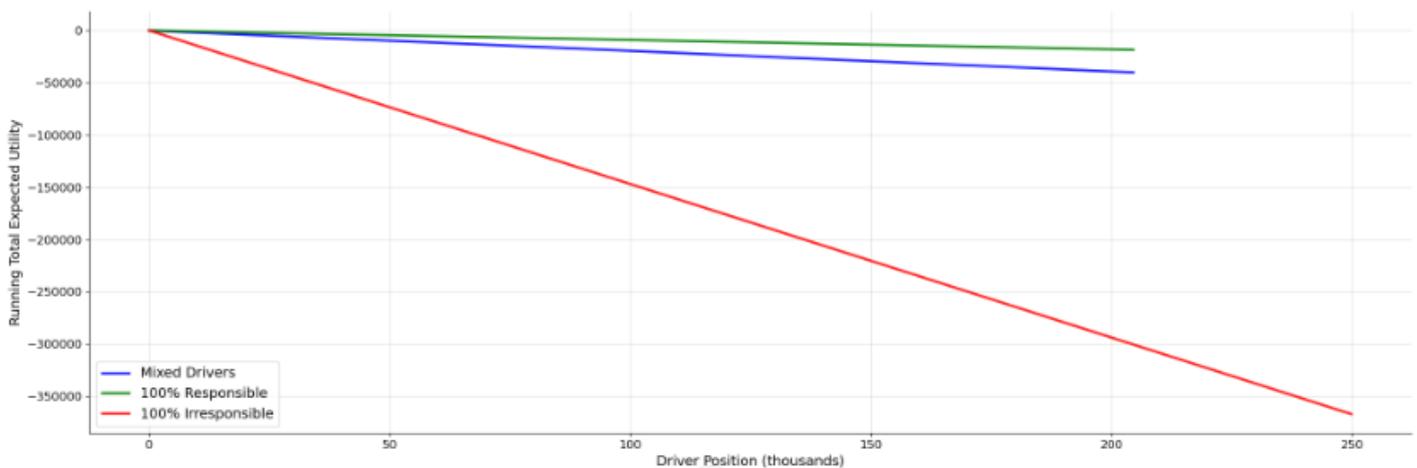

**Source: data analysis**

**Figure 4. Running total of expected driver utility for lane 2.**

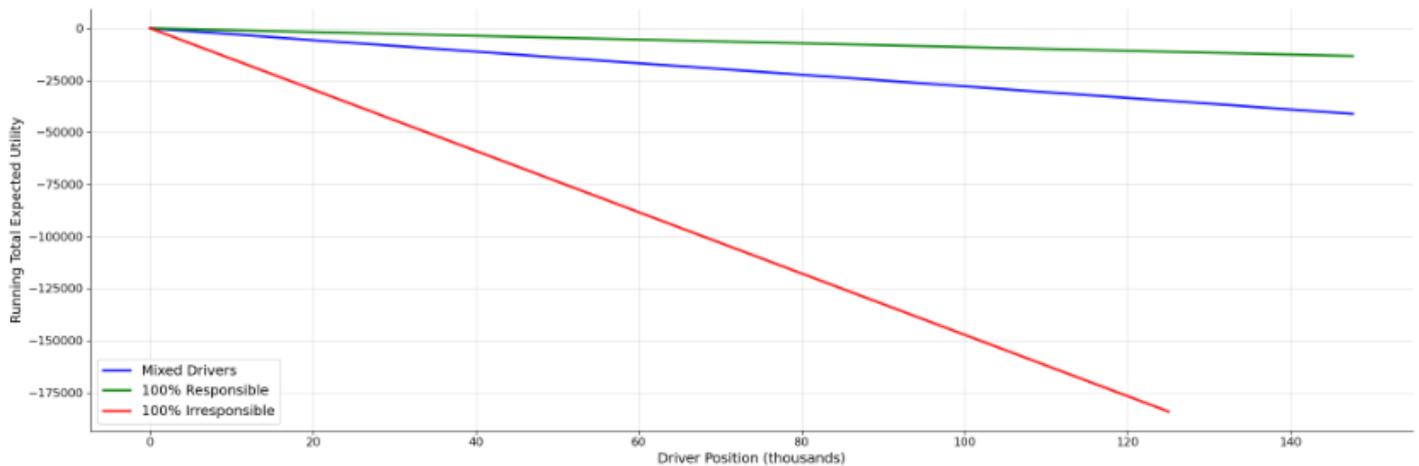

Source: data analysis

**Figure 5. Running total of expected driver utility for lane 3.**

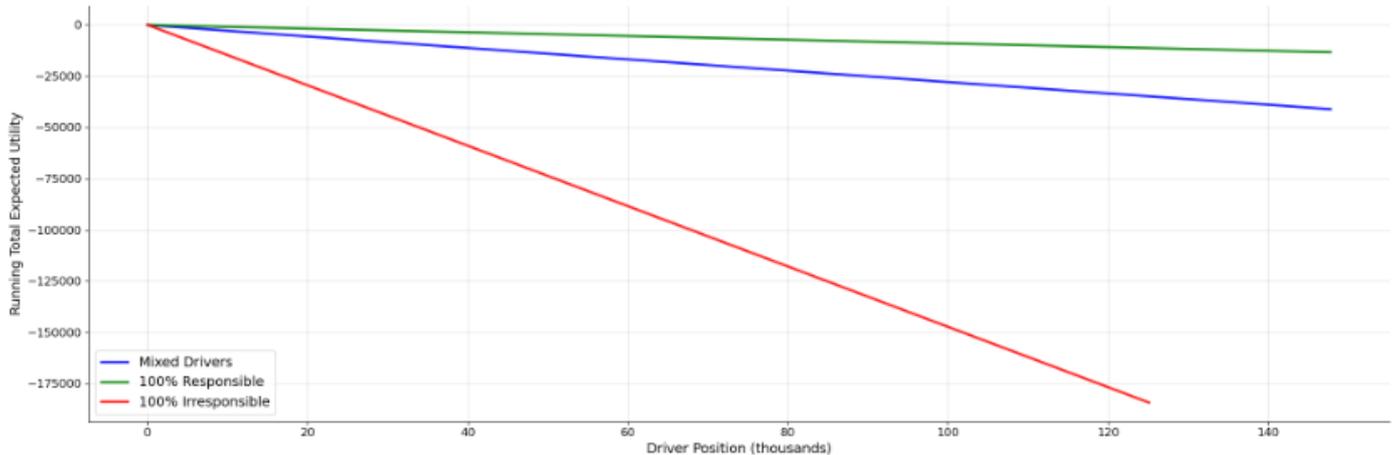

Source: data analysis

5. Discussion

The literature review identified several key themes. First, the diversity among drivers significantly influences the performance of traffic systems. Variations in risk tolerance, time valuation, and levels of cooperation among different driver types contribute to complex dynamics that frequently result in inefficiencies. Second, the strategic interactions among drivers give rise to game-theoretic scenarios, wherein individually rational actions culminate in collective irrationality, leading to traffic congestion that could be alleviated through improved coordination. The literature suggests that the formation of traffic jams is a multifaceted process shaped by driver behavior diversity, strategic interactions, equilibrium dynamics, and decision-making under uncertainty. Insights from behavioral economics, game theory, and mathematical modeling have expanded our understanding of traffic systems beyond the scope of traditional engineering methods.

The Driver behavior archetype model provides a comprehensive quantitative framework for understanding traffic flow dynamics by modeling individual driver behavior. The extensive simulation results revealed the following: 1. A heterogeneous mix of driver types results in a systematic reduction in utility. 2. Responsible driving is a disadvantage in diverse traffic environments. 3. Traffic congestion predictably emerges from patterns of driver interaction. 4. Improving the system as a whole requires coordinated changes in the distribution of driver behavior. The cumulative expected utility for drivers across all three lanes exhibits distinct patterns of utility decline that systematically vary based on driver composition and lane position (Figures ?-?). In scenarios with mixed drivers, lane 1 accommodated 204,531 drivers before system failure, whereas lanes 2 and 3 supported 147,641 and 147,828 drivers, respectively, underscoring the varying capacity of lane positions under diverse driver behaviors. The utility curves for mixed drivers demonstrate a typical pattern of accelerating degradation, where the decline becomes increasingly steep as more drivers enter the system, illustrating the nonlinear relationship between traffic density and individual utility observed in empirical traffic studies (Helbing, 2001; Treiber and Kesting, 2013). This accelerating decline indicates that each additional driver imposes a progressively severe marginal cost on the system, consistent with the theoretical predictions of congestion externalities in transportation economics (Vickrey, 1969; Arnott et al., 1993).

In contrast, the scenario in which all drivers exhibit responsible behavior results in markedly different utility patterns. Although the total number of drivers in lanes 1, 2, and 3 (204,531, 147,641, and 147,828, respectively) remains consistent with the mixed scenario, the utility curves assume a distinct configuration. Responsible drivers demonstrated a consistent, nearly linear decline across all lanes, indicating that cooperative driving behavior promotes a more stable and predictable relationship between traffic volume and individual utility. This linear decline suggests that responsible drivers distribute congestion costs more equitably among the driver population, thereby avoiding the severe utility collapse observed in other scenarios. The uniformity of this pattern across all three lanes supports the theoretical proposition that cooperative equilibria can sustain higher system efficiency, even as demand increases (Roughgarden, 2005; Wardrop, 1952).

Conversely, the all-irresponsible driver scenario revealed the most striking results, with utility curves exhibiting exponential-like degradation, characterized by sharp downward trends. Notably, Lane 1 in the irresponsible scenario accommodated significantly more drivers (249,859) than in the mixed or responsible scenarios (204,531), reflecting a 22.2% increase in throughput. However, Lanes 2 and 3 display the opposite trend, with the irresponsible scenario accommodating fewer drivers (125,011 and 125,130, respectively) compared to the mixed and responsible scenarios (approximately 147,000 each), indicating a 15.3% reduction in capacity. This uneven pattern suggests that aggressive driving behavior may temporarily enhance throughput in the main lane (Lane 1) through more aggressive merging and shorter following distances; however, this occurs at the expense of severe degradation in the secondary lanes and a rapid utility collapse. The steep slope of the irresponsible driver utility functions indicates that the system approaches breakdown much more rapidly, with each additional driver imposing exponentially increasing costs on all participants.

The diverse breakdown points across various scenarios and lanes provide significant insights into the impact of driver behavior on system resilience. The observation that lane 1 extends further under reckless driving conditions, while lanes 2 and 3 contract, suggests that aggressive driving results in an uneven distribution of traffic flow. This phenomenon can lead to the concentration of vehicles in the primary lane, thereby destabilizing the secondary lanes. This conclusion aligns with empirical studies on traffic instability and lane-specific breakdowns, as documented in traffic flow theory (Kerner, 2004; Treiber et al., 2000). The distinct patterns observed between lane types indicate that comprehending the system-level effects of driver behavior requires more than merely examining the overall throughput. The spatial distribution of traffic and lane-specific dynamics are essential for determining the overall system performance and the transition point from free-flowing to congested traffic conditions.

The results of the Mann-Whitney U tests reveal a distinct division in how different driver archetypes influence the expected utility within the traffic system. Statistical analysis demonstrates that driver archetypes fall into two separate categories with significantly different utility outcomes. The first category comprises irresponsible drivers, such as Dangerous, Slowpoke, Selfish, Left Lane Camper, and Impatient archetypes, who exhibit mean expected utilities between 0.147 and 0.149. There were no statistically significant differences among these subtypes, as all p-values exceeded 0.05, and the effect sizes were negligible. In stark contrast, the second category consists entirely of Responsible drivers, who have a mean expected utility of -0.102, which is substantially different from all irresponsible archetypes (Cohen's d ≈ 2.03, $p < 0.001$).The substantial effect size between the responsible and irresponsible driver groups has significant implications for understanding traffic dynamics. A Cohen's d value of approximately 2.0 is exceptionally large in statistical terms, particularly given that effect sizes over 0.8 are typically considered "large." An effect size of 2.0 indicates that the two groups are separated by two standard deviations, signifying not merely a quantitative difference but a fundamental structural divergence in how the game-theoretic system treats these archetypes. This statistical separation provides empirical evidence that the payoff structure within traffic interactions results in qualitatively different outcomes for cooperative and non-cooperative strategies. These findings provide robust evidence that responsible driving behavior yields consistently superior outcomes across all utility measures, whereas irresponsible driving incurs systematic penalties. The model successfully differentiates and rewards cooperative behavior, demonstrating its effectiveness in modeling driver behavior dynamics within traffic systems.

An examination of total lane utility reveals a paradoxical relationship between driver composition and system-level outcomes. With the current composition of 40.9% responsible drivers and 59.1% irresponsible drivers, the total lane utility was 9,355.75. However, this aggregate figure conceals a notable asymmetry in contributions: responsible drivers collectively contribute -8,543.5 to total utility, representing a significant negative impact on the system, while irresponsible drivers contribute +17,899.25. This finding is particularly striking because despite constituting only 40.9% of the driver population, responsible drivers create a negative utility sink that nearly offsets the positive contributions of all irresponsible drivers combined. This asymmetry suggests that the current equilibrium is maintained not through balanced contributions but through the exploitation of cooperative behavior by non-cooperative actors.

From a game theory perspective, these findings illuminate several key dynamics that help explain the persistence of traffic congestion. First, responsible drivers act as "utility absorbers" in the system, shouldering a disproportionate share of the costs associated with traffic jams, while irresponsible drivers reap the benefits of aggressive driving and non-cooperative tactics. Second, the significant difference in effect size between the groups suggests that being "responsible" is a dominated strategy in game-theoretic terms—rational individuals aiming to maximize their expected utility would shift to irresponsible behavior. This results in Nash equilibrium instability, where the choice that is rational for individuals (shifting to irresponsible driving) conflicts with the outcome that is optimal for everyone (universal cooperation). Third, the system exhibits classic "Tragedy of the Commons' dynamics, where individual incentives promote irresponsible behavior, even though widespread adoption of such strategies leads to worse collective outcomes, such as traffic jams and decreased overall efficiency. The nearly perfect symmetry in utility transfers, with irresponsible drivers gaining approximately 0.148 and responsible drivers losing approximately 0.102, indicates that the system primarily operates through zero-sum dynamics rather than creating or destroying value. Utility appears to be transferred from responsible to irresponsible drivers rather than being generated anew, suggesting that the current traffic system functions as a redistributive mechanism that rewards defection and penalizes cooperation.

The results support the main argument of the model that when individuals prioritize their own interests through reckless driving, it inherently opposes the collective benefit achieved by driving responsibly, and the system penalizes those who choose to cooperate. The data indicate that conventional enforcement strategies aimed at increasing the number of responsible drivers would ironically decrease the overall lane efficiency by increasing the number of individuals who consume resources without altering the fundamental reward system that encourages noncooperation. This suggests that successful policy measures should concentrate on altering the framework of the system itself—using methods such as dynamic pricing, redesigning infrastructure, and aligning incentives–rather than attempting to change the strategies of participants within the current system. The application of the model across three parallel lanes revealed significant spatial dynamics in the traffic flow. In Lane 1, typically the leftmost or fastest lane, 204,531 drivers were accommodated in both the mixed and responsible scenarios, whereas 249,859 drivers were observed in the irresponsible scenario. This suggests that aggressive driving behavior can temporarily enhance throughput in high-speed lanes by reducing the following distances and facilitating more aggressive merging. Conversely, Lanes 2 and 3 exhibited a different pattern, with the irresponsible scenario accommodating only approximately 125,000 drivers compared to roughly 147,000 in the other scenarios. This discrepancy indicates that the driver behavior archetype model functions differently depending on the lane position, with aggressive behavior causing an uneven traffic distribution that concentrates vehicles in lane 1 while destabilizing the secondary lanes. The traffic jam condition of the model ($U\_total \leq 0$) is reached at varying points in different lanes, which accounts for the different breakdown points in the running total utility curves across scenarios and lane positions. The corollaries of the model provide further insight into the simulation outcomes. Corollary 1.1(i) states that total utility is capped by the sum of baseline utilities, implying that strategic interactions and type mismatches can only decrease utility from its theoretical maximum. This clarifies why even the all-Responsible scenario experiences declining utility as more drivers join, as strategic interactions in congested conditions incur costs, even among cooperative drivers. Corollary 1.1(ii) verifies that homogeneous driver populations (all-Responsible or all-Irresponsible) eliminate adjustment penalties, resulting in linear and exponential patterns observed in the utility curves. Corollary 1.1(iii) quantifies the cost of heterogeneity at precisely 0.5 utility units per type mismatch, enabling accurate calculation of the total penalty caused by driver diversity in the mixed scenario. These properties collectively elucidate why the mixed scenario leads to the characteristic accelerating degradation pattern: each additional driver not only contributes to their own strategic interaction costs, but also increases the likelihood of type mismatches throughout the lane.

This framework provides a robust foundation for the advancement of policy formulation to potentially modify driver behavior through various means. By quantitatively correlating individual driver behavior with system-level outcomes, it enables the assessment of traffic management interventions, prediction of congestion patterns based on behavioral distributions, and optimization of lane configurations and control systems. Furthermore, it facilitates the design of incentive mechanisms that encourage responsible driving, thereby aligning individual incentives with collective efficiency and enabling the development of targeted strategies for congestion mitigation.

## 6. Directions for future research

The findings of this analysis suggest several promising avenues for future research that could significantly enhance our understanding of traffic dynamics and driver–behavior interactions. First, developing mechanisms for dynamically adjusting payoff matrices based on real-time traffic density would facilitate more accurate modeling of how driver utilities and strategic interactions evolve as congestion levels vary throughout the day. This approach would enable the model to capture the nonlinear relationships between traffic volume and individual decision-making processes. Second, expanding the current framework to incorporate multi-lane interaction modeling would provide deeper insights into how drivers decide to change lanes and how these decisions propagate through the traffic system. This would particularly emphasize the strategic interdependencies between drivers in adjacent lanes and the equilibrium patterns that emerge from these complex spatial interactions. Third, a critical next step involves rigorous integration with real-world traffic data validation, necessitating the collection of empirical data from actual highway systems using sensors, cameras, and connected vehicle technologies. This would assess the model's predictive accuracy and adjust parameters to reflect the observed driver behavior patterns in various geographic and cultural contexts. Finally, developing targeted intervention strategies to enhance the overall system utility could leverage the theoretical insights from this research to design practical policy measures. These could include dynamic pricing mechanisms, real-time information systems, or behavioral nudges that promote cooperation and mitigate the negative externalities of selfish driving behaviors, ultimately translating game-theoretic principles into actionable traffic management solutions that benefit both individual drivers and transportation systems.

## 7. Limitations

Although this analysis provides valuable insights into the strategic interactions among drivers and the development of traffic congestion patterns, it is essential to acknowledge several key limitations. First, the model employs a simplified two-action choice framework, wherein drivers can only choose to follow or pass. This inherently restricts the complex range of real-world driving behaviors, such as lane changes, speed adjustments, merging maneuvers, and adaptive responses to surrounding traffic (Kesting et al., 2007; Treiber & Kesting, 2013). Although this binary decision-making structure is analytically manageable, it may not adequately capture the continuous nature of driver decision-making and the nuanced differences between aggressive and conservative driving styles observed in empirical studies (Brackstone & McDonald, 1999; Toledo, 2007). Second, the reliance on static payoff matrices presents a significant limitation, as these matrices do not dynamically adjust to reflect changing traffic conditions, such as increasing congestion density, varying speeds, or the mix of different driver types in the traffic flow (Nagel & Schreckenberg, 1992; Helbing, 2001). In reality, the utilities associated with specific driving actions are highly context-dependent and can change significantly as traffic shifts from free-flow to congested states, indicating that future models should incorporate adaptive payoff structures that respond to real-time traffic parameters (Chowdhury et al., 2000; Maerivoet and De Moor, 2005). Third, the current framework offers limited consideration of external environmental factors that significantly impact driver behavior and traffic dynamics, including adverse weather conditions such as rain, snow, or fog that reduce visibility and road friction; physical road characteristics such as grade, curvature, and surface quality; and temporal factors such as time of day, day of the week, and seasonal variations in traffic patterns (Maze et al., 2006; Rakha et al., 2012). Beyond these model-specific constraints, it is crucial to recognize the inherent limitations of game-theoretic approaches to traffic analysis more broadly. Game theory assumes rational decision making by all agents, yet extensive behavioral research has shown that human drivers often exhibit bounded rationality, emotional responses, and cognitive biases that deviate from optimal strategic behavior (Kahneman & Tversky, 1979; Simon, 1955). Additionally, classical game theory often struggles to adequately represent the spatial and temporal dynamics inherent in traffic systems, where interactions occur continuously across space and time, rather than in discrete, well-defined games (Helbing & Tilch, 1998). The assumption of complete information—that drivers have perfect knowledge of other drivers' types, intentions, and payoffs—is particularly problematic in traffic contexts where information is inherently incomplete and asymmetric (Harsanyi, 1967; Fudenberg & Tirole, 1991). Furthermore, identifying and maintaining the stability of Nash equilibria in large-scale traffic systems with diverse driver populations remains computationally challenging and may not always produce unique or meaningful solutions (Roughgarden, 2005; Wardrop, 1952). These limitations collectively suggest that while game-theoretic models provide valuable theoretical frameworks for understanding strategic driver interactions, they must be complemented with empirical validation, behavioral insights, and more sophisticated modeling techniques that can capture the full complexity of real-world traffic systems.